\documentclass[twocolumn,showpacs,footinbib,bibnotes,prl,floatfix]{revtex4}
\usepackage{graphicx}
\usepackage{amsmath}
\usepackage{amssymb}

\begin{document}

\title{Precision measurement of light shifts in a single trapped Ba$^+$ ion}
\author{J. A. Sherman}\email{jeff.sherman@gmail.com}
\author{T. W. Koerber}\thanks{Now with Institut f\"{u}r Experimentalphysik, Universit\"{a}t Innsbruck}
\author{A. Markhotok}
\author{W. Nagourney}
\author{E. N. Fortson}
\affiliation{Department of Physics, Box 351560, University of
Washington, Seattle, WA 98195-1560}
\date{\today}

\newcommand{\Sstate}{\ensuremath{6S_{1/2}} }
\newcommand{\Pstate}{\ensuremath{6P_{1/2}} }
\newcommand{\Pstatespecial}{\ensuremath{6P_{1/2,3/2}} }
\newcommand{\Pupper}{\ensuremath{6P_{3/2}} }
\newcommand{\Dstate}{\ensuremath{5D_{3/2}} }
\newcommand{\Dshelve}{\ensuremath{5D_{5/2}} }
\newcommand{\Fstate}{\ensuremath{4F_{5/2}} }
\newcommand{\Ba}{\ensuremath{^{138}\text{Ba}^+} }
\newcommand{\Baodd}{\ensuremath{^{137}\text{Ba}^+} }

\newcommand{\lsrResult}{-11.494(13)}
\newcommand{\lsrResultSeparateErrors}{-11.494(08)(10)}
\newcommand{\argonWavelength}{514.531}

\newcommand{\jSectionHeading}[1]{}

\newcommand{\jfrac}[2]{#1/#2}

\begin{abstract}
Using a single trapped barium ion we have developed an rf spectroscopy technique to measure the ratio of the off-resonant vector ac Stark effect (or light shift) in the \Sstate and \Dstate states to 0.1\% precision.  We find $R \equiv \Delta_S / \Delta_D = \lsrResult$ at \argonWavelength~nm where $\Delta_{S,D}$ are the light shifts of the $m = \pm 1/2$ splittings due to circularly polarized light.  Comparison of this result with an {\it ab initio} calculation of $R$ would yield a new test of atomic theory.  By appropriately choosing an off-resonant light shift wavelength one can emphasize the contribution of one or a few dipole matrix elements and precisely determine their values.
\end{abstract}


\maketitle

\jSectionHeading{Introduction}
Experimental tests of atomic theory often involve the measurement of atomic state lifetimes, oscillator strengths, polarizabilities~\cite{Snow05}, and other properties which depend directly on atomic dipole matrix elements.  Absolute measurements of these quantities can be difficult.  Another approach \cite{Sieradzan04} is to make high precision measurements of properties which can be directly calculated using modern atomic theory techniques and depend on ratios of atomic matrix elements.  Here we report a 0.1\% measurement of the ratio $R$ of the ac Stark effect (or light shift) in the \Sstate and \Dstate states of a singly-ionized barium ion, iso-electronic to the well-studied alkali atom Cs.  Comparison of this result with an {\it ab initio} calculation of $R$ would yield a new test of atomic theory.

Since $R$ is expressible as ratios of matrix elements (shown below), this measurement also establishes a sum rule relating the barium matrix elements known to $\sim 1\%$ or better (i.e.\ $\langle \Sstate || r || \Pstatespecial \rangle$) to matrix elements with about ten times worse experimental precision \cite{Guet91, Kastberg93, Dzuba01, Gopakumar02}.  One of the latter, $\langle \Dstate || r || \Pstate \rangle$, is crucial for a proposed atomic parity violation experiment \cite{Fortson93,Safronova99,Koerber03}.  In our scheme, a particular matrix element can be studied by choosing a light shift wavelength close to the corresponding dipole transition so that other contributions remain small.  The technique is generalizable to other trapped ion species with metastable states.

\begin{figure}
    \includegraphics[width=\columnwidth]{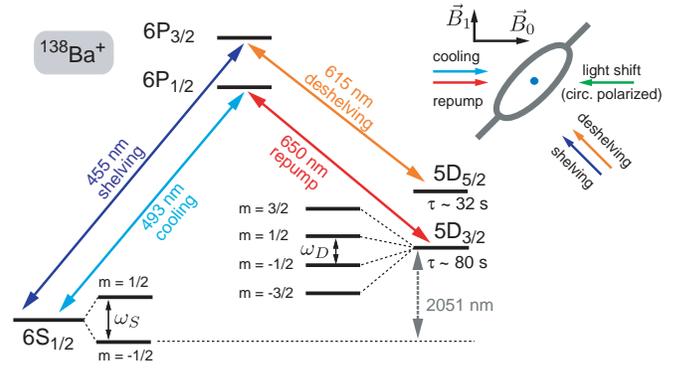}
     \caption[barium energy diagram]
{This energy level diagram of \Ba shows the optical transitions and magnetic spin resonances $\omega_S$ and $\omega_D$ of interest in this work as well as the electric-dipole forbidden 2051~nm transition important to a proposed atomic parity violation experiment.  Note the several second lifetimes \cite{Yu97,Madej90} of the metastable \Dstate and \Dshelve states.  The inset cartoon of a Paul-Straubel~\cite{Yu91} trap and ion shows the alignment of the lasers and rf spin-flip field $\vec{B}_1$ with respect to the static magnetic field $\vec{B}_0$.  The \Fstate state, while relevant to the measurement, is not shown.}
    \label{fig:levels}
\end{figure}

\jSectionHeading{Light shift theory}
To second order in perturbation theory an off-resonant light beam of intensity $I$, frequency $\omega$, and polarization $\boldsymbol{\hat{\epsilon}}$ causes a frequency shift $\Delta_{k,m}$ in atomic level $| k,m \rangle$ of:
\begin{equation}
\label{eq:ls}
\Delta_{k,m} = 2 \pi \alpha I \sum_{k',m',\pm \omega} \frac{ \left| \langle k,m | \boldsymbol{\hat{\epsilon}} \cdot \mathbf{r} | k',m' \rangle \right|^2}{ W_{k'} - W_{k} \pm \hbar \omega},
\end{equation}
where $\alpha$ is the fine-structure constant, $k$ stands for atomic quantum numbers $nlj$, and $W_k$ is the unperturbed energy of level $k$.  Though it is difficult to measure the light intensity at the site of a trapped ion, $I$ cancels in a measured ratio of light shifts.  The light shift due to circularly polarized off-resonant light acts effectively as a magnetic field pointing along the direction of light propagation \cite{Mathur68, Cohen-Tannoudji72}.  In our case, a laboratory magnetic field ($B_0 \sim$ 2.5 G) is aligned with the light shift beam. Therefore, if the same magnetic sublevels in two atomic states are used, there is only second order dependence of the ratio on polarization errors and misalignment of the light shift beam with respect to the magnetic field.

By employing an rf spectroscopy technique \cite{Koerber02} that features the metastable \Dshelve ``shelving'' state as a means of detecting spin flips, we make measurements of the $m=-\jfrac{1}{2} \leftrightarrow +\jfrac{1}{2}$ splittings $\omega_S$ and $\omega_D$ in the \Sstate and \Dstate states of a single \Ba ion without any light shift, which we interleave with measurements of the splittings, $\omega_S^{LS}$ and $\omega_D^{LS}$, made with application of the light shift beam.  The measured ratio of light shifts $R$ is then
\begin{align}
R &\equiv \frac{\Delta_S}{\Delta_D} = \frac{\omega_S^{LS} - \omega_S}{\omega_D^{LS} - \omega_{D}} \label{eq:lsrmeas} \\ \intertext{which is expressed using Eq.~(\ref{eq:ls}) as}
 R  &= \frac{\displaystyle \sum_{k', \pm \omega} f_{s}(k,k') \frac{ | \langle \Sstate || r || k' \rangle |^2}{(W_{k'} - W_{\Sstate} \pm \hbar \omega)}}{ \displaystyle \sum_{k', \pm \omega} f_{d}(k,k')  \frac{| \langle \Dstate || r || k' \rangle |^2}{(W_{k'} - W_{\Dstate} \pm \hbar \omega)}}.  \label{eq:lsr}
\end{align}
Here $f_s$ and $f_d$ are known angular factors.  The sums extend over continuum states.  Core effects and higher-order terms are significant at or above the $\sim 1\%$ level and must be included in a precision atomic calculation.

\begin{figure}
        \includegraphics[width=\columnwidth]{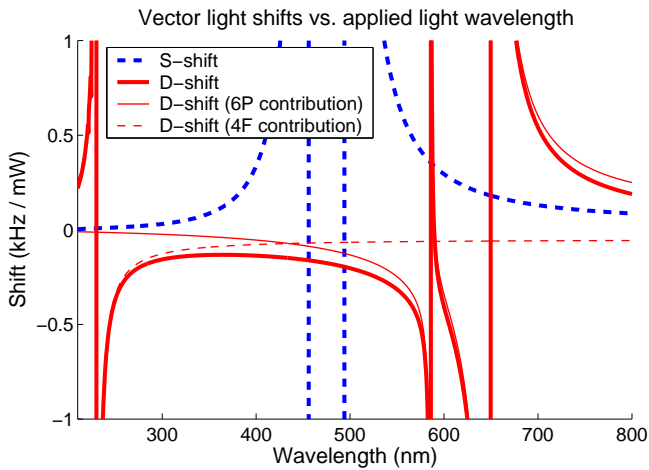}
    \caption[LSContrib]
{Estimated light shifts of the \Sstate (thick dashed) and \Dstate (thick solid) $m= -\jfrac{1}{2} \leftrightarrow \jfrac{1}{2}$ Zeeman resonances as a function of applied light shift wavelength assuming a 20~$\mu$m spot.  Note the divergences near atomic resonances.  Separately plotted are the $\langle \Dstate || r || \Pstatespecial \rangle$ contributions to the shift in the \Dstate state (thin solid) which dominates in the visible spectrum and the $\langle \Dstate || r || \Fstate \rangle$ contribution (thin dashed).  The calculation includes electric dipole coupling with the $6P_{1/2}$, $6P_{3/2}$, $7P_{1/2}$, $7P_{3/2}$ \cite{Guet91, Kastberg93, Dzuba01, Gopakumar02} and $4F_{5/2}$ \cite{SahooPrivate} states.}
    \label{fig:lsContrib}
\end{figure}

Because the light shift of level $k$ due to level $k'$ depends inversely on the detuning between the applied light frequency and the splitting between $k$ and $k'$, choosing a light shift frequency $\omega$ close to an allowed dipole resonance will make one or a few terms of the sum in Eq.~(\ref{eq:lsr}) dominate.  For instance, the contributions of the $\langle \Sstate || r || \Pstatespecial \rangle$ and $\langle \Dstate || r || \Pstatespecial \rangle$ terms dominate the shifts $\Delta_S$ and $\Delta_D$ for most visible wavelengths as shown in Figure~\ref{fig:lsContrib}.

\jSectionHeading{Method}
We trap a single \Ba ion by applying several hundred volts of 10~MHz rf to a twisted wire Paul-Straubel~\cite{Yu91} trap.  We cool the ion to the Lamb-Dicke regime on the 493~nm $\Sstate \leftrightarrow \Pstate$ transition using a frequency doubled 986~nm diode laser while also applying a 650~nm beam which prevents pumping to the long-lived \Dstate state.  This repump laser is frequency modulated at 2~kHz to a depth of $\sim$100~MHz to frequency lock it opto-galvanically to a barium hollow cathode lamp and to eliminate coherent effects which otherwise appear on the cooling transition \cite{Raab98}.  A static magnetic field aligned along both lasers splits the magnetic sublevels by a few MHz and eliminates dark states \cite{Berkeland02}.  High intensity filtered light-emitting diodes perform shelving to and deshelving from the \Dshelve state with rates $\gtrsim$10~Hz.

\begin{figure}
        \includegraphics[width=\columnwidth]{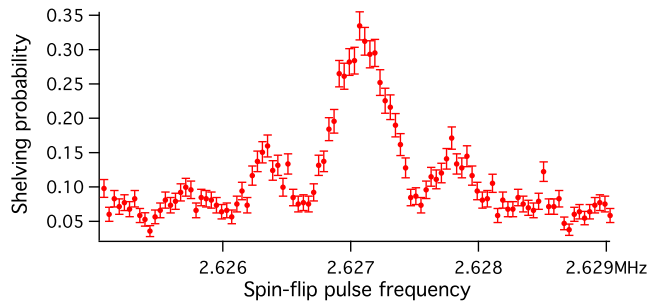}
\caption[Data 1]
{Example of a \Dstate resonance scan.  The probability that the ion is ``shelved'' in the metastable \Dshelve state is correlated to whether an rf $\pi$-pulse was on resonance with the \Dstate $m=-\jfrac{1}{2} \leftrightarrow +\jfrac{1}{2}$ splitting.  Here, homogenous broadening dominates the line width and coherent features unique to this four state system are visible.}
    \label{fig:data1}
\end{figure}

Because earlier reports \cite{Koerber02, Koerber03} gave details of the rf spectroscopy measurement in the \Sstate state and since an important systematic effect arises from the \Dstate resonance line shape, we describe the measurement in the \Dstate state in detail here.  The red repumping beam is aligned with the magnetic field.  Making it left circular polarized ($\sigma^-$) with an electro-optic modulator causes optical pumping into a state in the lower manifold ($m=-\jfrac{1}{2}, -\jfrac{3}{2}$) of $\Dstate$.  With the cooling and repumping beams turned off, a properly timed rf pulse resonant with the $m=-\jfrac{1}{2} \leftrightarrow +\jfrac{1}{2}$ splitting moves the ion to the upper ($m=+\jfrac{1}{2}, +\jfrac{3}{2}$) manifold with moderate probability.  We must now read out the occurrence of this rf spin-flip.  Applying a pulse of $\sigma^-$ repumping light to cause a decay to $\Sstate$ followed by a pulse of shelving (455~nm, $\Sstate \to \Pupper$) light transfers an ion in the upper \Dstate manifold to the \Dshelve shelving state while leaving the lower manifold population untouched.  Now detection of 493~nm fluorescence upon reapplication of the cooling and repumping beams is correlated to whether the ion is in the \Dshelve shelved state and hence whether the rf pulse was on resonance.  By repeating the measurement to gather statistics at several spin-flip radio frequencies one obtains resonance line shapes as shown in Figure~\ref{fig:data1}.

The size and shape of the coherent features (e.g.\ sidebands) of the line depend primarily on the initial condition of the ion; that is, with what probability $a$ it is prepared via optical pumping into the $m=-\jfrac{1}{2}$ state instead of $m=-\jfrac{3}{2}$.  In principle $a \le 0.8$ because of differential pumping rates to $\Pstate$; lower values will result from laser/magnetic field misalignment and poor red laser polarization quality.  By fitting resonances using the optical Bloch equations we have determined that $0.5 \le a \le 0.78$ are typically observed values.

We chose the convenient \argonWavelength~nm (air wavelength) line of an argon-ion laser as the off-resonant light shifting beam.  Approximately 1~W of laser light travels through an acousto-optic modulator (AOM) used for intensity stabilization and couples into a single-mode polarization-maintaining fiber after which are circular-polarizing optics and the intensity stabilization photodiode.  The use of the AOM, fiber, polarization optics, and sensor in this order translates laser pointing and polarization noise into intensity noise which is nulled by the AOM servo to much better than $0.1\%$ over a dc-10~kHz bandwidth.

\jSectionHeading{Data} 
A single data run consists of four interleaved measurements: \Sstate and \Dstate Zeeman resonances with and without the application of a light shift beam.  The four resonances are typically limited by inhomogeneous broadening due to small fluctuations in the magnetic field or light shift beam intensity, pointing, and polarization. They are therefore fit to Gaussians to determine the resonance centers and uncertainties.  We assign statistical errors to each light shift ratio measurement by propagating these uncertainties through Eq.~(\ref{eq:lsrmeas}).  About two hours is required to reach a statistical precision of $1 \%$ per measurement.

\begin{figure}
    \includegraphics[width=\columnwidth]{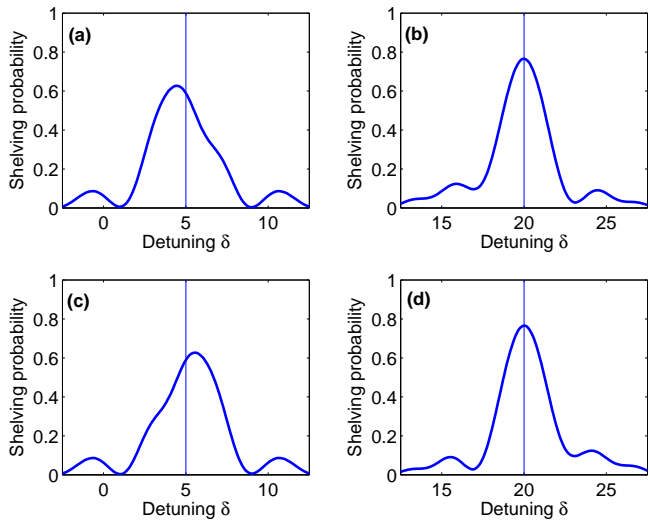}
\caption[Data 2]
{The most important systematic effect arises from distortions in the \Dstate resonance line shape caused by the tensor piece of the light shift.  The vertical lines indicate the simulated light shift values $\Delta_D$ in units of $\Omega_D$.  Shown in (a) and (c) are simulated line shapes for $\Delta_D / \Omega_D = 5$ while (b) and (d) show $\Delta_D / \Omega_D = 20$.  The initial conditions for (a) and (b) were $a = 0.78$ into the lower spin state manifold of \Dstate while (c) and (d) were prepared in the upper spin state manifold.}
    \label{fig:lineshape}
\end{figure}

\begin{figure}
     \includegraphics[width=\columnwidth]{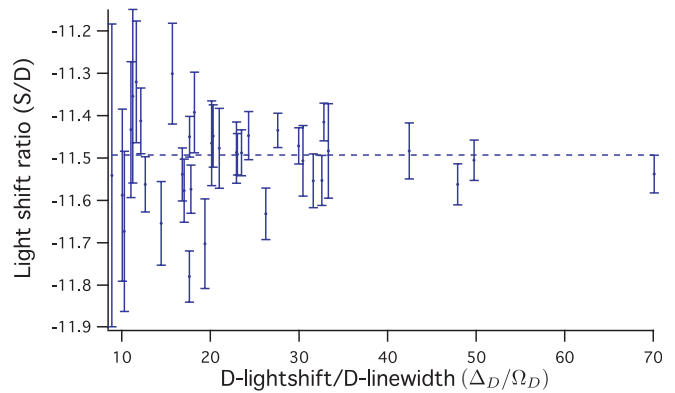}
\caption[Data 3]
{Observed light shift ratios $R$ defined in Eq.~(\ref{eq:lsrmeas}) are shown against the most significant systematic parameter: the ratio of measured light shift in the $\Dstate$ state and the spin-flip Rabi-frequency $\Delta_D/ \Omega_D$.  For low values of this ratio, the tensor light shift distorts the D-resonance line shape leading to systematic shifts of either sign and variable size.  If only data with $\Delta_D/ \Omega_D > 20$ are considered (putting any potential systematic below $0.1 \%$ in our models) the straight line fit $R = \lsrResult$ explains the data with $\chi_\text{r} = 1.0$.  
}
    \label{fig:data3}
\end{figure}

\jSectionHeading{Systematics}
Systematic effects common to shifted and unshifted resonance measurements completely cancel.  Many other systematic effects are eliminated because we interleave resonance measurements in both the \Sstate and \Dstate states and are only concerned with the ratio of their shifts.  Effects that remain are discussed below; they fall into the categories of line shape effects, alignment and polarization errors, ion displacements, magnetic field variations, coupling to the trapping rf field, and spectral purity of the light shift laser.

The most significant systematic effect is caused by the line shape of the light shifted \Dstate resonance.  $J > 1/2$ in this state, so a tensor component to the light shift makes the outer splittings $m = -\jfrac{3}{2} \leftrightarrow -\jfrac{1}{2}$ and $m = +\jfrac{1}{2} \leftrightarrow +\jfrac{3}{2}$ different from the splitting of interest $m = -\jfrac{1}{2} \leftrightarrow +\jfrac{1}{2}$.  This has been observed to cause distortions in the \Dstate resonance line shapes when the light shifts $\Delta_D$ are comparable to the spin-flip Rabi oscillation frequency $\Omega_D$.  The distortion near the center of the line decreases markedly with increasing $\Delta_D / \Omega_D$ as illustrated in Figure~\ref{fig:lineshape}.  The distortion is worse at higher $a$ values and at light shift wavelengths with significant $\langle \Dstate || r || \Fstate \rangle$ contribution.  The sign of the line-center error changes with reversing the light shift beam polarization relative to the magnetic field, or with initially preparing the ion in the opposite $\Dstate$ manifold.  We confirmed these reversals, which explain the increased scatter seen in data with low $\Delta_D / \Omega_D$ (see Figure~\ref{fig:data3}) since the sign of the effect was not controlled.  We minimize this systematic by only including data in the final analysis for which $\Delta_D \gg \Omega_D$.  By choosing the cut-off $\Delta_D / \Omega_D > 20$ our models suggest the average remaining effect is less than $0.05\%$.

A misalignment $\alpha$ of the light shift beam from the magnetic field decreases observed light shifts and alters the measured $R$.  Throughout the experiment the light shifts $\Delta_S$ and $\Delta_D$ are kept small with respect to the splittings $\omega_{S}$ and $\omega_D$ so that these errors are quadratic in $\alpha$ and multiplied by factors $\Delta_S/\omega_S$ or $\Delta_D/\omega_D$.  For a typical data run, $\Delta_S \sim 100$~kHz at $\omega_S \sim 6.8$~MHz, we estimate an $\alpha = 10^\circ$ misalignment causes $<0.1 \%$ change in $R$ while independently we have constrained $\alpha < 5^\circ$.  Errors in the light shift laser polarization couple to misalignments to produce systematic shifts in the \Dstate splitting due to the tensor light shift. These are kept well below $0.1 \%$ by maintaining a measured polarization quality of $\sigma > 0.95$ in addition to the above misalignment bounds.

The pseudo-potential due to the rf trap completely overpowers the expected dipole force experienced by an ion in a gradient of the light shift beam ($P<100$~mW, with a $\sim 20$~$\mu$m spot).  We have ruled out any effects which might displace the ion in a beam gradient by taking some data with the ion near the edge of the beam to set bounds and taking the majority of data with the ion centered in the beam spot to $\pm 5 \%$. We detected no dependence of $\Delta_S$ on the Rabi spin-flip frequency $\Omega_S$ which would imply displacement of the ion caused by the spin-flip rf.  Changes in the magnetic field at the site of the ion accompanying the application of the light shift laser, for instance due to the activation of a shutter, have been ruled out at the 1~Hz level by running the experiment with the light shift laser turned off and seeing no anomalous shift.

Capacitive pickup of the trap rf creates magnetic fields that shift (like the Bloch-Siegert effect) the $\Sstate$ and $\Dstate$ resonances up to a few kHz.  This effect is calibrated for each run by comparing the measured ratio of $g$-factors $\omega_S / \omega_D$ against its well measured value \cite{Knoll96}.  Most of this effect is common to both shifted and unshifted resonances since $\Delta_S \ll \omega_S$ and $\Delta_D \ll \omega_D$ by design; we apply corrections, no higher than $0.06 \%$ in any single run, to account for the net effect on $R$.

Commercial wavemeters confirmed no significant  wavelength drift with occasional small changes in argon-ion laser tube pressure.  Background argon-ion fluorescence coupled via the optical fiber to the ion was measured below $20 \mu$W which can only account for sub-Hz light shifts.

\jSectionHeading{Conclusions and future} 
All data are shown in Figure~\ref{fig:data3}.  By cutting data with $\Delta_D / \Omega_D < 20$ to minimize systematics as discussed above we obtain the result $R = \lsrResult$, where the quoted uncertainty includes a statistical contribution of $\sigma_\text{stat.} = 0.008$ and total systematic uncertainty $\sigma_\text{syst.} = 0.010$ added in quadrature.  This agrees with a theoretical estimate of $R_\text{est.} = 11(1)$ obtained using published matrix elements and uncertainties (see Figure~\ref{fig:lsContrib} note).  An \emph{ab initio} calculation of $R$ would provide an exacting test of atomic theory.
 
In conclusion, we have demonstrated that the measurement of off-resonant light shifts in single trapped ions via electron shelving is a viable technique for precisely testing atomic theory.  Future plans include performing this measurement again at a different off-resonant wavelength in order to untangle $\langle \Dstate || r || \Pstatespecial \rangle$ and $\langle \Dstate || r || \Fstate \rangle$ contributions in Eq.~(\ref{eq:lsr}).  The apparatus might also be used to measure the \Dstate electronic quadrupole moment via a strong dc potential applied to the trap ring~\cite{Yu94}.  Such quadrupole Stark shifts impact single ion frequency standards \cite{Itano00} and merit further study in a calculable system such as $\Ba$.  


\jSectionHeading{Acknowledgments}
This work is supported by NSF grant numbers PHY-0099535.  The authors thank W.\ Johnson and M.\ S.\ Safronova for their discussion of atomic theory issues.

\bibliography{lspaper}

\end{document}